\documentclass[sigconf]{acmart}  

\AtBeginDocument{%
  \providecommand\BibTeX{{%
    \normalfont B\kern-0.5em{\scshape i\kern-0.25em b}\kern-0.8em\TeX}}}

\setcopyright{acmcopyright}
\copyrightyear{2018}
\acmYear{2018}
\acmDOI{XXXXXXX.XXXXXXX}

\acmConference[Conference acronym 'XX]{Make sure to enter the correct
  conference title from your rights confirmation emai}{June 03--05,
  2018}{Woodstock, NY}

\acmPrice{15.00}
\acmISBN{978-1-4503-XXXX-X/18/06}

\begin{document}

\title{Robust Interaction-Based Relevance Modeling for Online E-Commerce Search}

\author{Ben Chen}
\email{chenben.cb@alibaba-inc.com}
\orcid{0000-0003-4495-8686}
\affiliation{%
  \institution{Alibaba Group}
  \streetaddress{Wangshang Road 699}
  \city{Hangzhou}
  \state{Zhejiang}
  \country{China}
  \postcode{430074}
}

\author{Huangyu Dai}
\email{daihuangyu.dhy@alibaba-inc.com}
\orcid{0000-0002-3844-8359}
\affiliation{%
  \institution{Alibaba Group}
  \streetaddress{Wangshang Road 699}
  \city{Hangzhou}
  \state{Zhejiang}
  \country{China}
  \postcode{430074}
}

\author{Xiang Ma}
\email{qianxiang.mx@alibaba-inc.com}
\affiliation{%
  \institution{Alibaba Group}
  \streetaddress{Wangshang Road 699}
  \city{Hangzhou}
  \state{Zhejiang}
  \country{China}
  \postcode{430074}
}

\author{Wen Jiang}
\email{jiangwen.jw@alibaba-inc.com}
\affiliation{%
  \institution{Alibaba Group}
  \streetaddress{Wangshang Road 699}
  \city{Hangzhou}
  \state{Zhejiang}
  \country{China}
  \postcode{430074}
}

\author{Wei Ning}
\email{wei.ningw@alibaba-inc.com}
\affiliation{%
  \institution{Alibaba Group}
  \streetaddress{699 Wang Shang Road}
  \city{Hangzhou}
  \state{Zhejiang}
  \country{China}
  \postcode{43017-6221}
}

\renewcommand{\shortauthors}{Ben Chen, et al.}

\begin{abstract}
Semantic relevance calculation is crucial for e-commerce search engines, as it ensures that the items selected closely align with customer intent. Inadequate attention to this aspect can detrimentally affect user experience and engagement. Traditional text-matching techniques are prevalent but often fail to capture the nuances of search intent accurately, so neural networks now have become a preferred solution to processing such complex text matching. Existing methods predominantly employ representation-based architectures, which strike a balance between high traffic capacity and low latency. However, they exhibit significant shortcomings in generalization and robustness when compared to interaction-based architectures. In this work, we introduce a robust interaction-based modeling paradigm to address these shortcomings. It encompasses 1) a dynamic length representation scheme for expedited inference, 2) a professional terms recognition method to identify subjects and core attributes from complex sentence structures, and 3) a contrastive adversarial training protocol to bolster the model's robustness and matching capabilities. Extensive offline evaluations demonstrate the superior robustness and effectiveness of our approach, and online A/B testing confirms its ability to improve relevance in the same exposure position, resulting in more clicks and conversions. To the best of our knowledge, this method is the first interaction-based approach for large e-commerce search relevance calculation. Notably, we have deployed it for the entire search traffic on alibaba.com, the largest B2B e-commerce platform in the world.
\end{abstract}







\maketitle

\section{Introduction}
In online e-commerce platforms, the search engine's effectiveness hinges on two core capabilities: identifying users' explicit demands through queries and mining purchasing preferences from historical click logs. A comprehensive search solution requires not only a ranking module that enhances click-through and conversion rates but also a relevance module that validates the appropriateness of displayed items. \textbf{S}emantic \textbf{R}elevance \textbf{C}alculation (SRC), a fundamental component of e-commerce platforms, discerns core keywords within short queries against long item descriptions to accurately score and rank pertinent items \cite{chen2023beyond, chen2015convolutional, hu2019survey}. Prioritizing co-click conversion modeling without considering users' intent and matching relevance can erode user attention and, consequently, engagement and conversion rates over time.

Semantic relevance calculation can be classified as a domain-specific text matching task, markedly distinct from general tasks such as MS MARCO \cite{nguyen2016ms} and STS \cite{cer2017semantic}, which measure semantic similarity in standard language contexts, or semantic question answering (matching) \cite{liu2018lcqmc, dhakal2018exploring, antoniou2022survey}, which concentrate on the primary themes of queries and documents. SRC for online e-commerce search addresses distinct challenges:

1) Query Intent and Keyword Clarity. SRC must distinguish concise user queries that often carry vague meanings and match them with the most matching items. A query like 'new apple discount' could ambiguously refer to a promotion on fresh produce, a reduced price on Apple electronics, or a clothing brand's latest offer. Moreover, product descriptions are frequently stuffed with extraneous keywords to gain more exposure, such as a dress described as "elegant evening gown summer crystal luxury sequin red cocktail party", which dilutes the significance of essential keywords and muddles the search accuracy.

2) Latency vs. Precision Trade-off. As e-commerce platforms strive for efficiency, the shift from traditional keyword-based search algorithms to advanced neural-based models marks a significant progression \cite{robertson1995okapi, xiao2019weakly, yao2022reprbert}. These neural models are divided into two types: representation-based and interaction-based. Representation-based models, leveraging siamese network architectures, encode queries and items into compact embeddings efficiently, making them suitable for high-traffic online searches according to their computational speed. However, their oversimplification often leads to poorer relevance predictions. In contrast, interaction-based models excel in capturing subtle semantic relationships, offering finer distinction and accuracy, but the intensive computational demands limit their practical application in real-time search scenarios with stringent latency requirements.

3) Enhancing Robustness and Generalization. The diverse linguistic expressions arising from cultural differences complicate the accuracy of SRC models, such as various terms for the same discounts of "50\% off sale", "half price promotion", and "discounted by half". Meanwhile, to reduce computational overhead, both representation-based and interaction-based methods often use techniques like pruning or distillation \cite{hinton2015distilling, jiao2019tinybert, wang2020minilm} to simplify models for efficient real-time processing. However, these condensed models, while effective on familiar data, struggle with unfamiliar datasets, revealing limitations in their robustness and generalization. An efficient solution is to expand the training data diversity, but manual annotation is resource-intensive with limited scope, and noisy historical log data can dilute the models' effectiveness. These further weaken the generalization ability of SRC.

In summary, while the conventional interaction-based model BERT can achieve state-of-the-art performances on real query-item search logs, its computational intensity remains a barrier to its direct implementation in online search engines, despite efforts to mitigate this through distillation and pruning. Consequently, the less computationally demanding representation-based models are often the default choice. To address this challenge, in this paper we introduce a robust interaction-based method for relevance modeling. It encompasses three key innovations. The first is a dynamic-length representation scheme. It can intelligently scale input token size to match the varying lengths of queries and item descriptions, thereby optimizing computational resources. 

The second is an efficient professional terms recognition strategy. It enhances the model's vocabulary with industry-specific phrases and employs Named Entity Recognition (NER) to highlight subjects and core attributes, so as to reinforce the representation of professional terminology. Finally, to counteract the performance limitations of shallow models, we devised a contrastive adversarial training (CAT) mechanism. It can bolster the model's generalization and robustness by simultaneously optimizing the embedding representations of both inputs and outputs. Impressively, this optimized model, with just 3 layers, outperforms traditional 12-layer BERT base models in efficiency and effectiveness.

We conducted comprehensive offline evaluations using annotated query-item relevance pairs derived from user search logs, and the results showcased significant performance enhancements, affirming the efficacy and robustness of our proposed method. Online A/B testing results also showcased that it can improve the matching relevance of query-item in the same exposure position, and attract more clicks and conversions. To the best of our knowledge, it is the first interaction-based relevance calculation work for a large-scale e-commerce search engine, accommodating the daily needs of tens of millions of users and serving billions of retrieval page views (PV). Moreover, this method has been successfully deployed for the entire search traffic on alibaba.com, the world's largest B2B e-commerce platform, and has yielded substantial improvements in conversion rates across the board.

The contributions of this paper are summarized as follows:
\begin{itemize}
\item We propose a robust interaction-based method for modeling query-item semantic relevance, which can handle thousands of candidates for a given query simultaneously, serving tens of millions of users with billions of retrieval page-views everyday. To the best of our knowledge, this is the pioneering work of applying interaction-based relevance computation to such a large-scale e-commerce search engine.
\item We introduce one dynamic length query/item representation scheme to reduce the inference time, one professional terms recognition strategy to identify subject and core attributes, as well as a contrastive adversarial training mechanism to enhance the relevance matching performance of the interaction-based model.
\item This approach has been successfully deployed on the entire search traffic for alibaba.com, the world's largest B2B e-commerce platform, for over 12 months. It has contributed to a rise in click-through and conversion rates, thereby significantly enhancing industry revenue through various iterations.
\end{itemize}

The remainder of the paper is organized as follows: Section 2 introduces the related works for SRC, and Section 3 details the proposed methods. Experimental results and analysis are presented in Section 4. Conclusions are drawn in Section 5.

\section{related works}
Semantic relevance calculation, fundamentally a text-matching task, has benefitted from a wide array of techniques. These range from traditional methods like TF-IDF and BM25 \cite{robertson1995okapi, manning2009introduction} to machine learning approaches such as DSSM \cite{huang2013learning}, LSTMs \cite{palangi2014semantic, wan2016match, palangi2016deep}, and CNNs \cite{chen2015convolutional, pang2016text, shen2014latent}. Despite their contributions, these models have limitations in bridging the vocabulary gap and accurately identifying key terms.
With the rapid development of BERT-based models enhancing performance across NLP tasks \cite{vaswani2017attention, devlin2018bert, liu2019roberta, lan2019albert, he2020deberta, sun2021ernie}, some advanced implementations like Sentence-BERT, BERT-flow, and SimCSE \cite{reimers2019sentence, li2020sentence, su2021whitening, gao2021simcse} have emerged as the preferred methods for complex text processing. However, these architectures still struggle with capturing the subtleties of search intent and are generally resource-intensive. Consequently, tailored strategies are necessary to overcome these specific challenges.

\begin{figure*}[htbp]
  \centering
  \includegraphics[width=\linewidth]{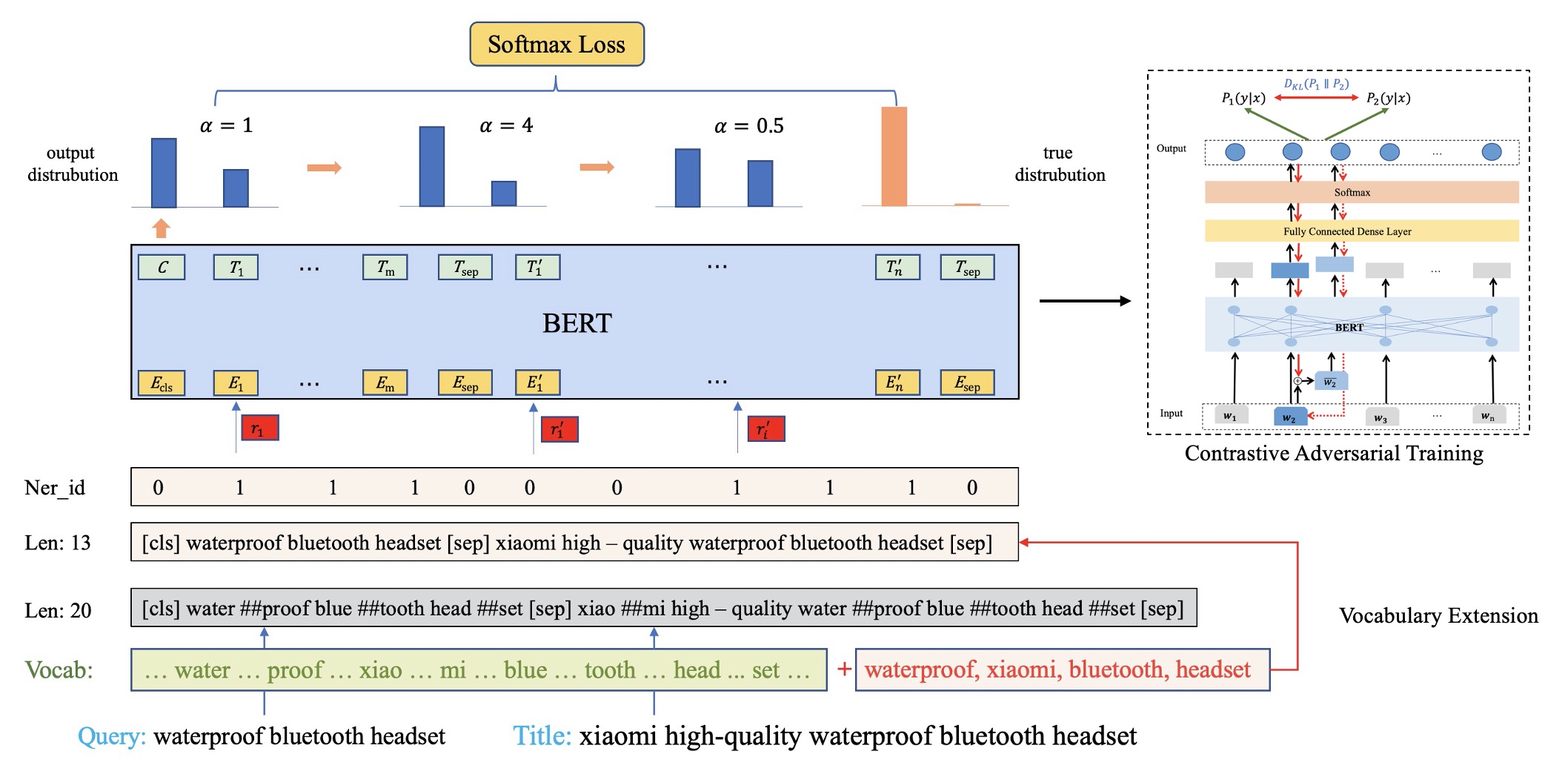}
  \caption{The overview of the ei-SRC, the proposed e-commerce interaction-based semantic relevance calculation method. For brevity, the dynamic-length representation scheme is not illustrated, which will be displayed separately.}
  \label{fig:figure1}
\end{figure*}

A common paradigm utilizes a softmax function on the final [CLS] token or average pooling outputs which are then scaled to a range between 0 and 1 to represent the likelihood of relevance \cite{nogueira2019passage, chen2023beyond}. Nogueira et al. implemented a multi-stage BERT-based architecture for ranking, incorporating innovative point-wise and pair-wise classification strategies \cite{nogueira2019multi}. Wu et al. introduced a multi-task learning framework aimed at minimizing the query-item vocabulary gap while optimizing multiple objectives \cite{wu2022multi}. Jiang et al. developed a data-driven relevance prediction framework by distilling knowledge from BERT and its sophisticated teacher models \cite{jiang2020bert2dnn}. Garakani et al. utilized a cross-encoder BERT model for query-item relevance prediction, further applying it to re-ranking and the optimization of search quality \cite{garakani2022improving}. While these methodologies surpass the accuracy of representation-based models, the vast traffic volume and the stringent latency requirements of live search environments present significant deployment challenges.
To address this, ReprBERT \cite{yao2022reprbert} proposes a unique solution that distills the interaction-based capabilities of BERT into a more streamlined representation-based model, employing dual interactive strategies to refine latent semantic interactions. This results in superior performance relative to conventional representation-based models. Moreover, the very recent Interactor \cite{ye2022fast} can capture fine-grained phrase-level information with a flexible contextualized interaction paradigm, and adopts a novel partial attention scheme to reduce the computational cost while maintaining the effectiveness. Nonetheless, their representational ability does not match that of fully interactive methods, so that the SRC performance does not quite reach the benchmark set by fully interaction-based models.

\section{methodology}
We use \textbf{ei-SRC} to indicate the proposed e-commerce interaction-based semantic relevance calculation method for brevity, and decompose its methodology into three components. Firstly, we outline two interaction strategies designed to minimize online computational overhead. Subsequently, we describe techniques developed to enhance the model's proficiency in handling domain-specific terminology. Lastly, we propose a novel training mechanism aimed at fortifying the model's representational capacity and accuracy in relevance matching. The comprehensive architecture of the ei-SRC method is depicted in Figure 1.

\subsection{Dynamic-length Representation Scheme}
The primary distinctions between representation- and interaction-based methodologies in SRC are rooted in their respective processing of queries and the associated computational frameworks. Representation-based techniques typically convert a query into a fixed-dimension vector, which simplifies queries to a uniform length. By pre-computing item embeddings offline and employing scoring functions like dot-product or cosine similarity online, these methods significantly streamline computational overheads, with the most resource-intensive step being the preliminary query processing. Conversely, interaction-based methods exhibit a higher sensitivity to query length. They operate by processing the query in real time, which involves grammatical normalization and tokenization to generate a sequence of tokens. These tokens are then merged with pre-processed item description tokens, culminating in an interactive and dynamic computation sequence. Ultimately, a non-linear classifier is utilized to determine the relevance score, providing a detailed assessment of the relationship between the query and the item \cite{nogueira2019passage}.


To address the varied requirements of industrial search applications and simultaneously evaluate thousands of item candidates per query, interaction-based methods typically pre-define a fixed token length for both queries and item descriptions. This standardization ensures uniform input sizes and consistent computational time across all pairings. However, this token length is often set longer to account for the occasional lengthy query or item description, leading to sub-optimal computational resource use for processing shorter texts. Such inefficiency impedes the real-world deployment of interaction-based methods in an online search platform.

Prior to introducing our optimization strategy, it is essential to explore the factors influencing the computational load of interaction-based models. 
The BERT framework is known for its multi-layered structure, where each layer principally consists of two core modules: the multi-head attention (MHA) and feed-forward neural network (FFN) sub-layers. Here we indicate $n$ as the batch size during training or the number of item candidates during online inference, $l$ as the total length of the input token sequence, and $d$ as the dimension of the embeddings. The number of attention heads is represented by $m$, and the attention head is set as $a$, such that $d=ma$.

The time complexity of multi-head attention sub-layers is:
\begin{equation}
  T_{MHA} = O(n*l^2*m^2*a) = O(n*l^2*d*m).
\end{equation}

The time complexity of FFN sub-layers is:
\begin{equation}
  T_{FFN} = O(n*l*k*m).
\end{equation}

\begin{figure}[tbp]
  \centering
  \includegraphics[width=\linewidth]{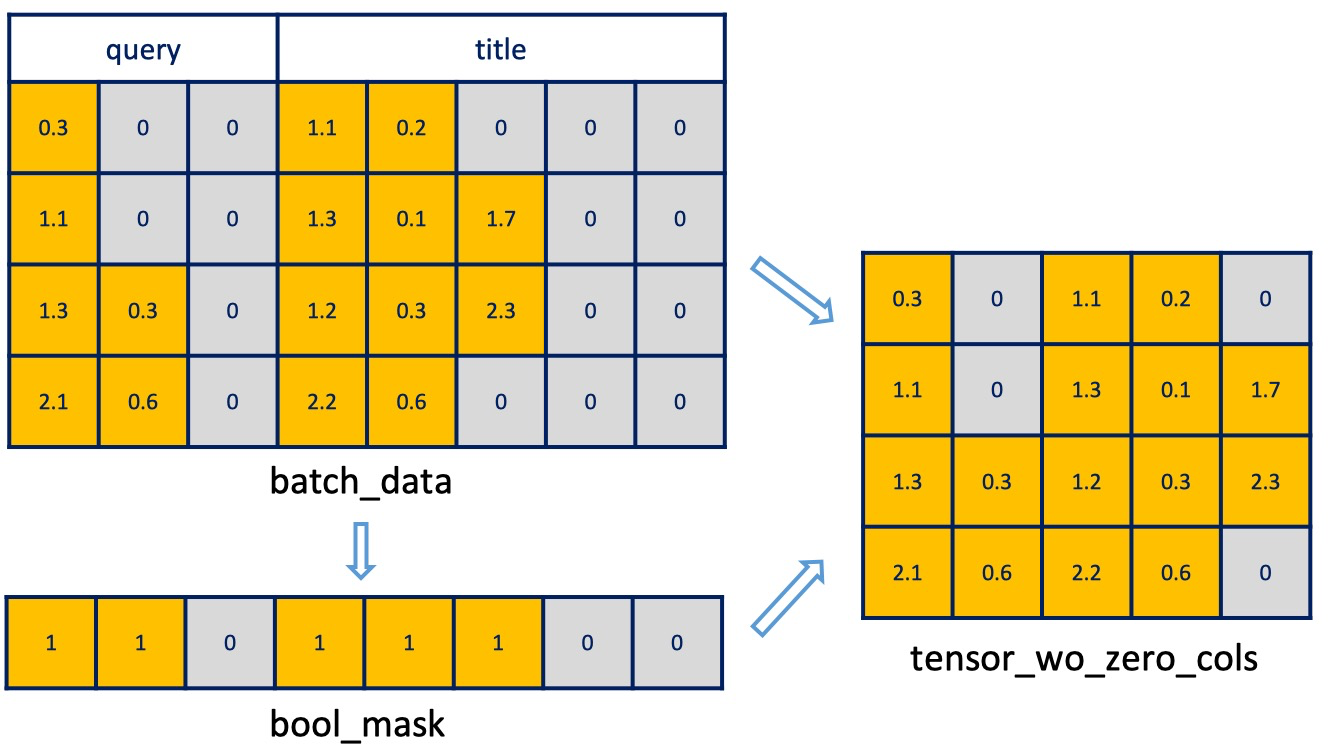}
  \caption{The overview of dynamic-length representation scheme, where each square is a token. The number shown on it simulates the sum of various types of embeddings.}
  \label{fig:figure2}
\end{figure}

\begin{table}
\begin{center}  
  \caption{Improvements in computational performance for each strategic combination employed within the \textbf{ei-SRC}}.
  \label{tab:comp_cost}
  \begin{tabular}{lccc}
    \toprule   
    Method & GPU utilization & reaction latency & AUC \\
    \midrule	  
    DRS & -34.61\% & -36.26\% & +0.4\% \\
    $\backslash$+Cache & -48.08\% & -40.66\% & - \\
    $\backslash$+Vocab. & -53.84\% & -46.15\% & +1.5\% \\
    $\backslash$+CAT. & -51.25\% & -44.83\% & +6.3\% \\
   \bottomrule   
\end{tabular}
\end{center}  
\end{table}

As model distilling or pruning could minimize $k$, $n$, and $d$, we can see that $T_{MHA}$ is proportional to the square of $l$, and $T_{FFN}$ have a linear correlation with $l$. If $l$ can be decreased, the total time consumption will be largely reduced.

Here we design a dynamic-length representation scheme to address this problem. Considering that the descriptions of queries and items are shorter than the settings for most cases, we can shorten the tokens' length by batch-dropping zero padding columns. To describe it precisely, we indicate $l_q$ as the query token length set in advance, and $l_i$ as the length of title tokens, then $l = l_q + l_i$. As shown in Figure 2, we calculate the max non-zero token length of queries ($l_q^{\prime}$) and items titles ($l_t^{\prime}$) in batch, and cut the full zero padding columns. The size of new input series is $l^{\prime} = l_q^{\prime} + l_i^{\prime}$, and time complexity $T_{MHA}$ is reduced to the original $(l^{\prime}/l)^2$ times.

Furthermore, we also pre-compute the tokens in advance for queries that the user mostly entered, the relevance scores for high-frequency query-item pairs, and then store them in the online cache to reduce the computation cost. The proportion of pre-computed pairs varies with the actual search. For example, we can use the top $20\%$ data in the past six months and implement daily updates.

In the real online A/B tests on alibaba.com, this method significantly reduced the computational cost and reaction latency without attenuation of the accuracy. Model architecture and implementation configurations will be detailed in the experimental section. Now we only list the comparison of calculations consumption in Table~\ref{tab:comp_cost}, where "DRS" indicates "Dynamic-length Representation Scheme", and "Cache" means the pre-computed results stored in the online cache. As for the computational performance, we use the GPU utilization, and reaction latency as the metrics.

\begin{table}[tbp]   
\begin{center}   
\caption{Top 20 words mostly used in alibaba.com, they would be split into many sub-tokens with the original WPE.}  
\label{tab:mostly_words}
\begin{tabular}{|c|c|c|c|c|}   
\hline   rohs & waterproof & 100\% & oem & pvc \\   
\hline   fokison & smd & cnc & acrylic & polyester  \\ 
\hline   osc & bluetooth & capacitors & 12v & scooter  \\  
\hline   mink & diy & wifi & diode & hoodie  \\      
\hline   
\end{tabular}   
\end{center}   
\end{table}

\begin{table}[tbp]
\begin{center}  
  \caption{Average sub-tokens numbers calculation for query-items pairs with original WPE and extended vocabulary.}.
  \label{tab:token_number}
  \begin{tabular}{lccc}
    \toprule   
    Types & original word & original WPE & extended vocab. \\
    \midrule	  
    each word & 1.00 & 1.37 & 1.22 \\
    each title & 15.61 & 21.46 & 19.06 \\
    each pairs & 26.41 & 41.62 & 36.55 \\
   \bottomrule   
\end{tabular}
\end{center}  
\end{table}

We can see that DRS can significantly reduce GPU utilization and response latency. Thanks to the reduction of interference caused by all zero padding columns, semantic representation and matching of query-item pairs have been enhanced, with a 0.4\% increase in AUC. In addition, offline pre-computation and online caching of query tokens and relevance scores for high-frequency pairs can significantly reduce the overhead of these repeated searches. Compared to the original model, these two methods combined contribute to the reduction of GPU utilization by 48\%, and response latency by 40\%, thereby easing the final resource consumption. 

\subsection{Professional Terms Recognition Strategy}
A notable feature of the tokenization tools WPE \cite{devlin2018bert} and BPE \cite{liu2019roberta} is that they do not cause OOV (Out Of Vocabulary) problems. BERT-based models can effectively combine the split tokens of a single word to express the semantic information well if they have trained on large corpora containing that word multiple times. However, in the e-commerce domain, queries and titles often contain specialized terms that are infrequently found in general corpora. This disparity can result in a misinterpretation of their intended meanings due to standard tokenization methods. For example, the term 'bluetooth' is typically segmented into "blue" and "\#\#tooth", and a lack of adequate training can lead to errors in matching. Consequently, substituting 'bluetooth' with variants like 'blue tooth' or even 'black tooth' in queries and titles might not significantly alter the calculated relevance score in matching, thereby illustrating a potential pitfall in accurately capturing the essence of e-commerce vocabulary.

To address this problem, here we propose an efficient professional terms recognition strategy. It contains two steps: First is the vocabulary extension. we calculated the frequency of each word contained in queries and item descriptions on alibaba.com in the past 12 months, selecting the top 20,000 words which will be split into many sub-tokens with the former tokenization, and added them into the original WPE vocabulary of BERT. Then the model would be continued pre-trained with the new vocabulary.

The top 20 words are listed in Table~\ref{tab:mostly_words}, we can see that extended words are obviously related to the e-commerce field. Moreover, we calculated the average number of sub-tokens for query-item pairs with these two vocabularies separately. As depicted in Table~\ref{tab:token_number}, the overall number of each query-item pair is reduced from 41 to 36, and it achieves a computational saving of nearly 22.81\% with Equation (1). For real online testings, we also find it can reduce the GPU utilization and response latency significantly, and further improve the metric AUC by nearly 1.5\%, as shown in Table~\ref{tab:comp_cost}. That is, it can not only ease the inference burden but also contribute to the improvement of the model's representation for specific terms.

Secondly, we implement NER \cite{wang2020automated, wang2020structure} to recognize the object and core keywords. We use the numbers 1-5 to mark "material", "function", "usage", "specification", "style", and "core" keywords in the query-item pairs, and generate the corresponding NER embeddings, which combine the token$\backslash$segmentation$\backslash$position embeddings to from the new input embeddings.

\subsection{Contrastive Adversarial Training Mechanism}
In order to reduce the time-consuming of online relevance inference, pruning or distillation is often adopted to get the shallow-layer models, resulting in the degradation of the model's generalizations to untrained pairs. A useful tip is training these models with a large range of effective data. However, highly reliable annotated pairs are costly and have low coverage, and the data sampling from user behaviors is more noisy because of many unintentional clicks. These all limit the online application of interaction-based methods.

To tackle this problem, here we propose one novel contrastive adversarial training (CAT) mechanism. It aims to improve the robustness of both input and output representations simultaneously and make better discrimination on hard samples. 

The CAT method is detailed as follows. We present query-item pair as $<q_i, p_i>$ (combined as the input $x_i$) and the relevance label as $y_i\in \{0,1\}$. \textbf{$\theta$} denotes the parameters of the corresponding model, and the basic objective for the SRC model is to minimize the negative log-likelihood loss function as:
\begin{equation}
  L_{BCE} = -\frac{1}{N} \sum_{i=1}^{n} \log p(y_i | x_i, \textbf{$\theta$})).
\end{equation}

For the first step, we adopt adversarial training \cite{miyato2016adversarial, zhu2019freelb} to reduce the model's sensitivity to perturbed input embeddings, and then improve robustness to various original examples. Specifically, we first calculate a small adversarial perturbation \textbf{${r_{adv}}$} based on the back-propagated gradient value:
\begin{equation}
  r_{adv}^i = -\epsilon \textbf{g}/\left\|\textbf{g}\right\|_2\text{ %
  where $\textbf{g}=\nabla{x_i} \log p(y_i | x_i, \textbf{$\theta$})$ },
\end{equation}
where $\epsilon$ is the coefficient weight to control the size of perturbation, then add it to the original input \textbf{$x$} to form one new perturbed embedding \textbf{$x+r_{adv}$}, and finally design a new function to fit the relevance label:
\begin{equation}
  L_{ATE} = -\frac{1}{N} \sum_{i=1}^{n} \log p(y_i | x_i + r_{adv}^i, \textbf{$\theta$}).
\end{equation}.

On the other hand, aiming to enhance the output representation, we try to minimize the difference between two output distributions from the same model with different dropouts. This strategy is adopted for two reasons: one is to eliminate the nonnegligible inconsistency between training and inference caused by the randomness introduced by dropout \cite{hinton2012improving, wu2021r}, and the other is to enhance the output representation. In detail, we randomly discard some non-substantial words filtered out by NER, to construct a new query-item pair with substantially the same semantics for each inference. Then the bidirectional KL divergence of their output passing through the same model should be kept minimal.

However, the simple combination of these two methods would cause four inferences, leading to a significant increase in computational cost. Considering the first inference of adversarial training only to be conducted for the $r_{adv}$ generation without any parameter updating, here we design one optimized scheme. That is, we retain the back-propagation of $L_{BCE}$, and then minimize the bidirectional KL divergence of output distributions between $x$ and $x_{adv}$ as:
\begin{equation}
\begin{aligned}
  L_{ADV} = \frac{1}{2N} \sum_{i=1}^{n} & (D_{KL}( p(y_i | x_i, \textbf{$\theta$}) || p(y_i | x_i + r_{adv}^i, \textbf{$\theta$})) \\
  &+  D_{KL}( p(y_i | x_i + r_{adv}^i, \textbf{$\theta$}) || p(y_i | x_i, \textbf{$\theta$})) ).
\end{aligned}
\end{equation}.

So all the computations only need two inferences, and the total loss can be formulated as follows:
\begin{equation}
  L_{total} = \alpha_1 \cdot L_{BCE} + \alpha_2 \cdot L_{ATE} + \alpha_3 \cdot L_{ADV}
\end{equation}
where $\alpha_1$, $\alpha_2$ and $\alpha_3$ are the weighting parameters.

Moreover, in order to remind the model to emphasize the hard mining examples, we replace the original softmax function with heated-up softmax \cite{zhang2018heated}:
\begin{equation}
  p_{(m_i|\textbf{$x_i$}, \textbf{$\theta$})} = \frac{exp(\alpha*z_i)}{exp(\alpha*z_0) + exp(\alpha*z_1)},
\end{equation}
where $z_i$ is the output logit of $x_i$, $m$ is the binary category with $m_i \in {0, 1}$, and $\alpha$ denotes the temperature parameter. 

We start the training process with a large $\alpha$, enabling the model to concentrate on hard samples. Subsequently, we gradually reduce $\alpha$ to shift the model's attention to boundary samples, and eventually, we set a minimal value for fine-tuning the easy pairs. As depicted in Figure 1, $\alpha$ adjusts the distribution of the last layer output and enables the model to focus on more hard mining samples as the training goes on.

As a side note, prior work \cite{miao2021simple} proposed a similar contrastive representation adversarial learning for text classification. However, it only cares about the input representation by minimizing the difference of outputs with increasing $\epsilon$ in Equation (4), but our CAT mechanism is designed to regularize the model by optimizing both input and output representations. The offline experiments below show that it can improve not only the robustness to examples with keyword repetition and stacking but also enhance the generalization for query-items matching of multiple industries.

\section{experiments}
In this section, we conduct comprehensive evaluations on manually annotated query-item pairs offline and rigorous A/B online testings online to verify the feasibility of the proposed methods.

\textbf{Dataset.} We extracted the highly reliable user-clicked pairs from alibaba.com's search logs over the past year to facilitate the continue pre-training. The selection of pairs was subject to two constraints: 1) Each query had to comprise a minimum of two words, including at least one core keyword, to filter out queries with ambiguous semantics such as 'dress' or 'sports shoes,' which often reflect unclear user search intents and result in a scattered range of clicked items. 2) Pairs exhibiting deeper click behaviors were re-sampled to a certain number, predicated on the basic cognition that query reflects the user’s intent and their clicks signify varying degrees of interest in different products. For alibaba.com, the cross-border B2B platform, user click behaviors are categorized into five escalating levels: page-click $\leq$ add-to-cart $\leq$ contact-supplier $\leq$ order $\leq$ pay. Based on multiple experimental tests, here we set the re-sampled number for each level as (1,1,2,3,5). In total, 80 million query-item pairs were collected, with 70 million designated for continuing pre-training, and the remaining 10 million high-frequency ones were reserved for subsequent fine-tuning.

Additionally, we further gathered a dataset of 250,000 query-item pairs and marked them as relevant or irrelevant via manual annotation. The relevance judgments were based on three criteria: the subject, the presence of core keywords, and units of measurement such as minimum order quantity, delivery time, and size specifications. A pair was marked as relevant only if it satisfied all three conditions. Each pair was assessed by three individuals to mitigate subjective bias. To further ensure accuracy, an expert specializing in search business reviewed the collective judgments. Subsequently, 150,000 annotated pairs were utilized for additional fine-tuning of models, while the remaining 100,000 served as the evaluation set.

\textbf{Baselines.} For comprehensive evaluations, we utilize the widely-employed models BERT, RoBERTa, StructBERT \cite{wang2019structbert}, and the very recent domain-specific SRC model ReprBERT \cite{yao2022reprbert}, as well as Interactor \cite{ye2022fast}, all pre-trained on the same dataset as base large models. Subsequently, we distilled BERT into smaller variants (L3-H128-A4), denoted as BERT$_{mini}$. Additionally, we trained a Sentence-BERT (SBERT) \cite{reimers2019sentence} and its distilled counterpart (SBERT$_{mini}$) to compare the efficacy of representation-based models. All models were initially fine-tuned on 10 million real-world query-item pairs and subsequently on annotated datasets.

\textbf{Implementation details.} 
During the pre-training phase, the learning rate and batch size were configured to 5e-5 and 32 respectively. For fine-tuning, we adjusted the learning rate to 2e-5, increased the batch size to 1024, and set the weighting parameters  ${\alpha_{1,2,3}}$ as (0.5, 0.5, $0.01$). The maximum token lengths were established at 16 for queries and 36 for item descriptions, while the actual input token lengths varied according to the dynamic-length representation scheme. All evaluated models were fine-tuned within 3 epochs with early-stopping. We employed the batch negative sampling strategy for the initial fine-tuning stage on 10 million user-clicked query-item pairs, designating other items in the batch as negative examples. All experiments were conducted using Tensorflow 1.12 for both online and offline evaluations, with the code implementations being made available for reproducibility.

\subsection{\textbf{Offline evaluation.}} 
For evaluating SRC as a binary classification task, we utilized AUC, Micro/Macro F1 scores, and Spearman's and Pearson's correlation coefficients as evaluated metrics. As depicted in Table~\ref{tab:offline}, comparative results for the base models ($X_{base}$ with 12 layers) and the scaled-down mini models ($X_{mini}$ with 3 layers) are presented for both representation-based and interaction-based approaches. It was observed that, at comparable parameter scales, representation-based models lagged in performance, exhibiting an approximate 3.5\% decrease in AUC when compared to interaction-based models. Furthermore, the $X_{base}$ models outperformed their $X_{mini}$ counterparts, reflecting a more robust interaction and representational capacity in larger models. Nonetheless, the computational intensity of $X_{base}$ models pose limitations for their practical deployment.

\begin{table}[tbp]
  \caption{Comparison with representation and interaction-based methods on manual annotated data}
  \label{tab:offline}
  \begin{tabular}{lcccc}
    \toprule   
    Strategy & AUC & F1 & Spearmanr & Pearsonr \\
    \midrule	  
    $SBERT_{m}$ & 0.8184 & 0.61/0.71 & 0.5515 & 0.3934 \\
    $SBERT_{b}$ & 0.8554 & 0.60/0.76 & 0.6061 & 0.4751 \\
     \midrule	  
    $BERT_{m}$ & 0.8500 & 0.65/0.77 & 0.6147 & 0.5079 \\
    $BERT_{b}$ & 0.8907 & 0.72/0.78 & 0.6742 & 0.6360 \\
    $ReprBERT_{b}$ & 0.8923 & 0.73/0.80 & 0.6804 & 0.6389 \\
    $Interactor_{b}$ & 0.8926 & 0.73/0.81 & 0.6810 & 0.6387 \\
    $RoBERTa_{b}$ & 0.8964 & 0.78/0.82 & 0.6865 & 0.6926 \\
    $StructBERT_{b}$ & 0.9011 & 0.78/0.81 & 0.6947 & 0.7096 \\
     \midrule	  
    ei-SRC (3L)  & \textbf{0.9033} & \textbf{0.81/0.83} & \textbf{0.6984} & \textbf{0.7181} \\
   \bottomrule   
\end{tabular}
\end{table}

\begin{table}[tbp]
  \caption{Ablation experiment results of increment strategies}
  \label{tab:ablation}
  \begin{tabular}{lcccc}
    \toprule   
    Method & AUC & F1 & Spearmanr & Pearsonr \\
     \midrule	  
    $BERT_{m}$ & 0.8500 & 0.65/0.77 & 0.6147 & 0.5079 \\
    $\backslash$+DRS & 0.8533 & 0.66/0.77 & 0.6395 & 0.5290 \\
    $\backslash$+AT & 0.8852 & 0.75/0.80 & 0.6720 & 0.6404 \\
    $\backslash$+CAT & 0.9033 & 0.81/0.83 & 0.6984 & 0.7181 \\
   \bottomrule   
\end{tabular}
\end{table}

Notably, despite the parity in parameter magnitude, ei-SRC surpasses the performance of two mini models SBERT${mini}$ and BERT${mini}$, with improvements of 10.37\% and 6.27\% respectively. Moreover, ei-SRC demonstrates superior performance across all metrics, even when compared to larger representation-based model SBERT$_{base}$ and interaction-based models BERT, RoBERTa, StructBERT, and recent domain-specific model ReprBERT, as well as Interactor. Remarkably, the 3-layer configuration of our model attains better results than the conventional 12-layer models on real-world search query-item pair evaluations. These findings not only underscore the efficacy of the methods proposed but also highlight the model's robustness and its ability to generalize across a diverse range of industrial applications.

To validate the efficacy of proposed strategies, we carried out supplementary ablation studies. These included assessments of computational performance, as summarized in Table~\ref{tab:comp_cost}, and comparisons of evaluated metrics, as detailed in Table~\ref{tab:ablation}. We observed a marked decrease in computational cost with the incremental application of dynamic-length representation (DRS), which led to a reduction of 34.61\%. The introduction of the online cache strategy contributed to a further reduction of 48.08\%, while the expansion of the vocabulary (Vocab.) yielded a decrease of 53.84\% in computational overhead. On the other hand, the incorporation of contrastive adversarial training (CAT) resulted in increased processing time due to its requirement for dual inferences. Nevertheless, CAT significantly enhanced the performance across all metrics: AUC increased by 5.85\%, F1 scores by 22.72\%/7.79\%, Spearmanr by 9.21\%, and Pearsonr by 35.74\%. As illustrated in Table~\ref{tab:ablation}, the three strategies collectively improved the model's matching results to different degrees. In conclusion, the combined application of these methods can conserve nearly half of the computational resources while substantially enhancing the model's relevance-matching capabilities.

\subsection{\textbf{Online Experiments.}} 
We implemented the ei-SRC model within the online search engine of www.alibaba.com, the largest B2B e-commerce platform in the world, which serves tens of millions of users generating billions of page-views (PVs) daily. The SRC module is integrated into the final stage of the search ranking process. Initially, it evaluates all candidate items—typically between 3,000 to 5,000-and filters out those with low correlation scores to ensure that no more than 2,000 items are selected for ranking. Subsequently, for each query-item pair, the SRC score is combined with the ranking model's scores and other factors to determine the final rank, so that one pair with higher SRC score would be listed in more forward exposure position.

In detail, the initial online model was a variant of sBERT$_{mini}$, and we sequentially conducted three A/B testings: BERT$_{mini}$ with Dynamic-length Representation Scheme ($\backslash$+DRS), followed by the inclusion of Adversarial Training ($\backslash$+AT), and finally, the incorporation of Contrastive Adversarial Training  ($\backslash$+CAT). We assessed the impact of each strategy on click through rate (CTR), average conversion rate (CVR), and average earnings rate (PAY) by comparing performance metrics before and after each implementation. Notably, the improvements reported are incremental relative to the preceding iteration model rather than the original BERT$_{mini}$, and the statistical significance (P-value) of all results was below 0.05. As indicated in Table~\ref{tab:ab_test}, there were substantial increases in CTR, CVR, and PAY after the application of each strategy, with the CAT strategy demonstrating particularly notable enhancements: CTR improved by 1.54\%, CVR by 1.02\%, and PAY by 1.87\%.

\begin{table}[tbp]
  \caption{Online A/B testing results in www.alibaba.com, all statistical significance (P-value) are smaller than 0.05}
  \label{tab:ab_test}
  \begin{tabular}{lcccc}
    \toprule   
    Strategy & CTR & CVR & PAY & P-value \\
     \midrule	  
    $\backslash$+DRS & +0.93\% & +1.09\% & +0.76\% & 0.018 \\
    $\backslash$+AT & +0.68\% & +1.37\% & +1.66\% & 0.023\\
    $\backslash$+CAT & +1.54\% & +1.02\% & +1.87\% & 0.004 \\
   \bottomrule   
\end{tabular}
\end{table}

\begin{table}[tbp]
  \caption{Manual evaluation results for online relevance, all testing pairs are extracted in the same exposure position}
  \label{tab:man_rel_eval}
  \begin{tabular}{lccc}
    \toprule   
    Strategy & Good & Fair & Bad \\
     \midrule	  
    $\backslash$+DRS & +0.91\% & -0.93\% & -3.61\% \\
    $\backslash$+AT & +4.01\% & -21.67\% & -3.13\% \\
    $\backslash$+CAT & +4.77\% & -3.87\% & -8.93\% \\
   \bottomrule   
\end{tabular}
\end{table}

Furthermore, to ascertain the actual impact on online search relevance, we conducted additional manual evaluations. Following each update, we randomly selected queries and extracted 2000 query-item pairs from identical exposure positions, ensuring all other variables remained constant. We engaged experts to rate each pair as 'Good' (both subject and core keywords match), 'Fair' (only subject matches), or 'Bad' (subjects differ). The outcomes of these assessments are presented in Table~\ref{tab:man_rel_eval}. The results indicate that each implemented strategy incrementally improved semantic relevance at the same exposure position, with the incorporation of CAT yielding the most significant enhancement. In conclusion, our proposed method has been demonstrated to substantially improve the user search experience, leading to increased clicks and conversions, and ultimately boosting the industry revenue.

Additional evaluation results, annotated test cases, and the source codes for dynamic-length representation scheme and contrastive adversarial training mechanism have been made publicly accessible\footnote {https://github.com/benchen4395/ei-src4search} to facilitate further research. We intend to expand our investigation into the integration of the ei-SRC framework within ranking models and its application to recommendation systems and P4P advertising. Moreover, we plan to examine the potential of Large Language Models (LLMs), such as ChatGPT\footnote{https://chat.openai.com} and ChatGLM\cite{zeng2022glm}, in guiding the learning process for SRC.

\section{Conclusion}
In this study, we introduced a robust interaction-based method for modeling semantic relevance in online e-commerce search engines. This method incorporates a dynamic length representation scheme, a professional terms recognition strategy, and a contrastive adversarial training scheme to improve relevance matching. Extensive experiments on offline annotated query-item pairs and rigorous online A/B testings have verified its effectiveness for enhancing the search experience and boosting the industry revenue. Significantly, this approach has been stably operational on www.alibaba.com, handling its full search traffic for over 12 months, undergoing several iterative improvements.

\bibliographystyle{ACM-Reference-Format}
\bibliography{sample-authordraft}

\end{document}